\catcode`\@=11
\newif\if@fewtab\@fewtabtrue

{\count255=\time\divide\count255 by 60
\xdef\hourmin{\number\count255}
\multiply\count255 by-60\advance\count255 by\time
\xdef\hourmin{\hourmin:\ifnum\count255<10 0\fi\the\count255}}
\def\ps@draft{\let\@mkboth\@gobbletwo
    \def\@oddhead{}
    \def\@oddfoot
       {\hbox to 7 cm{$\scriptstyle Draft\ version:\ \draftdate$
       \hfil}\hskip -7cm\hfil\rm\thepage \hfil}
    \def\@evenhead{}\let\@evenfoot\@oddfoot}

\def\ceqno{\global\@fewtabfalse
    \ifcase\@eqcnt \def\@tempa{& & &}\or \def\@tempa{& &}
      \or \def\@tempa{&}
      \or\def\@tempa{}\fi\@tempa
{\rm(\theequation)}}

\def\aeqno#1{\global\@fewtabfalse
    \ifcase\@eqcnt \def\@tempa{& & &}\or \def\@tempa{& &}
      \or \def\@tempa{&}
      \or\def\@tempa{}\fi\@tempa
{\rm(\theequation,#1)}}

\def\label#1{\ifnum\draftcontrol=1
 \global\def\draftnote{$\scriptstyle #1$}\fi
 \@bsphack\if@filesw {\let\thepage\relax
   \def\protect{\noexpand\noexpand\noexpand}%
\xdef\@gtempa{\write\@auxout{\string
      \newlabel{#1}{{\@currentlabel}{\thepage}}}}}\@gtempa
   \if@nobreak \ifvmode\nobreak\fi\fi\fi
  \@esphack}

\def\alabel#1#2{\label{#1}\global\@fewtabfalse
    \ifcase\@eqcnt \def\@tempa{& & &}\or \def\@tempa{& &}
      \or \def\@tempa{&}
      \or\def\@tempa{}\fi\@tempa
{\hbox to 3cm{\phantom{\rm(\theequation,#2)}
\draftnote \hfil}\hskip -3cm {\rm(\theequation,#2)}}}

\def\clabel#1{\label{#1}\global\@fewtabfalse
    \ifcase\@eqcnt \def\@tempa{& & &}\or \def\@tempa{& &}
      \or \def\@tempa{&}
      \or\def\@tempa{}\fi\@tempa
{\hbox to 3cm{\phantom{\rm(\theequation)}
\draftnote \hfil}\hskip -3cm{\rm(\theequation)}}}

\def\eqnarray{\def\draftnote{{}}\global\@fewtabtrue
\stepcounter{equation}\let\@currentlabel=\theequation
\global\@eqnswtrue
\global\@eqcnt\z@\tabskip\@centering\let\\=\@eqncr
$$\halign to \displaywidth\bgroup\@eqnsel\hskip\@centering\@eqcnt\z@
  $\displaystyle\tabskip\z@{##}$&\global\@eqcnt\@ne
  \hskip 1\arraycolsep \hfil${##}$\hfil
  &\global\@eqcnt\tw@ \hskip 1\arraycolsep
$\displaystyle\tabskip\z@{##}$
\hfil  \tabskip\@centering&\global\@eqcnt\thr@@\llap{##}\tabskip\z@
\cr}

\def\endeqnarray{\@@eqncr\egroup
      \global\advance\c@equation\m@ne$$\global\@ignoretrue}

\def\@eqnnum{\hbox to 3cm{\phantom{\rm(\theequation)} \draftnote
                         \hfil}\hskip -3cm {\rm(\theequation)}}

\def\@@eqncr{\let\@tempa\relax
    \ifcase\@eqcnt \def\@tempa{& & &}\or \def\@tempa{& &}
      \or \def\@tempa{&}
      \or\def\@tempa{}
\fi\@tempa
\if@eqnsw
\if@fewtab\@eqnnum\fi
\stepcounter{equation}\fi\global
\@eqnswtrue\global\@eqcnt\z@\global\@fewtabtrue\cr}

\def\draftcite#1{\ifnum\draftcontrol=1#1\else{}\fi}

\def\@lbibitem[#1]#2{\item{}\hskip -3cm \hbox to 2cm
{\hfil$\scriptstyle\draftcite{#2}$}\hskip
1cm[\@biblabel{#1}]\if@filesw
     {\def\protect##1{\string ##1\space}\immediate
      \write\@auxout{\string\bibcite{#2}{#1}}}\fi\ignorespaces}

\def\@bibitem#1{\item\hskip -3cm \hbox to 2cm
{\hfil $\scriptstyle\draftcite{#1}$}\hskip 1cm
\if@filesw \immediate\write\@auxout
       {\string\bibcite{#1}{\the\value{\@listctr}}}\fi\ignorespaces}

\def\nsection#1{\section{#1}}

\font\tendl=msbm10  scaled \magstep1
\font\sevendl=msbm7 scaled \magstep1
\font\fivedl=msbm5 scaled \magstep1
\font\tengl=eufm10  scaled \magstep1
\font\sevengl=eufm7 scaled \magstep1
\font\fivegl=eufm5 scaled \magstep1

\newfam\dlfam \def\dl{\fam\dlfam\tendl} 
\textfont\dlfam=\tendl \scriptfont\dlfam=\sevendl
\scriptscriptfont\dlfam=\fivedl
\newfam\glfam  
\textfont\glfam=\tengl \scriptfont\glfam=\sevengl
\scriptscriptfont\glfam=\fivegl

\def\draftdate{\number\month/\number\day/\number\year\ \ \ \hourmin }

\global\def\draftcontrol{0}
\catcode`\@=12
\def\tilde{\widetilde}
\def\hat{\widehat}
\def\pref#1{(\ref{#1})}
\documentstyle[11pt]{article}

\def\theequation{\arabic{equation}} 

\newcommand{\be}{\begin{eqnarray}}
\newcommand{\en}{\end{eqnarray}\vs 0.5 cm}

\newcommand{\no}{\noindent}
\newcommand{\vs}{\vskip}
\newcommand{\hs}{\hspace}

\newcommand{\un}{\underline}
\newcommand{\NR}{{{\dl R}}}
\newcommand{\NC}{{{\dl C}}}
\newcommand{\NZ}{{{\dl Z}}}
\newcommand{\qq}{\begin{eqnarray}}
\newcommand{\de}{\bar\partial}
\newcommand{\da}{\partial}
\newcommand{\ee}{{\rm e}}

\newcommand{\qqq}{\end{eqnarray}}

\newcommand{\tr}{\hbox{tr}}

\newcommand{\CL}{{\cal L}}

\newcommand{\CO}{{\cal O}}
\newcommand{\CP}{{\cal P}}

\newcommand{\s}{\hspace{0.05cm}}
\newcommand{\m}{\hspace{0.025cm}}

\newcommand{\hf}{{_1\over^2}}

\begin{document}
\title{Elliptic Wess-Zumino-Witten Model from Elliptic
Chern-Simons Theory}
\author{\ \\Fernando Falceto \\ Depto. F\'{\i}sica
Te\'orica, Univ. Zaragoza, E-50009 Zaragoza, Spain
\\ \\Krzysztof Gaw\c{e}dzki \\ I.H.E.S., C.N.R.S.,
F-91440  Bures-sur-Yvette, France}
\date{ }
\maketitle

\vskip 0.3cm
\vskip 1 cm

\begin{abstract}
\vskip 0.3cm

\noindent This letter continues the program
\cite{Quadr}\cite{fgk}\cite{g}\cite{g1}
aimed at analysis of the scalar
product of states in the Chern-Simons theory. It treats the
elliptic case with group $SU_2$.
The formal scalar product is expressed
as a multiple finite dimensional integral
which, if convergent for every state, provides
the space of states with a Hilbert space structure.
The convergence is checked for states with a single Wilson line
where the integral expressions encode the Bethe-Ansatz solutions
of the Lam\'{e} equation. In relation to the Wess-Zumino-Witten
conformal field theory, the scalar product
renders unitary the Knizhnik-Zamolodchikov-Bernard
connection and gives a pairing between conformal
blocks used to obtain the genus one correlation functions.
\end{abstract}
\vs 2cm

\nsection{Introduction}
\vskip 0.5cm

A principal object of the conformal field theory are
conformal blocks. For the group $SU_2$ Wess-Zumino-Witten
(WZW) model of level $k$ and Virasoro central charge
$c_k={3k\over^{k+2}}$,
on the elliptic curve $\NC/2\pi(\NZ+\tau\NZ)$, the
conformal blocks are given by the expressions
\qq
\gamma_{\tau,\un z}(u)\ =\
{\rm tr}_{\hat{V}_{k,l_1}}\left(\m
\phi^{l_1}_{j_n\m l_N}(z_N)
\ \phi^{l_N}_{j_{n-1}\m l_{N-1}}(z_{N-1})\ \cdots\
\phi^{l_2}_{j_1\m l_1}(z_1)\s\ q^{L_0-c_k/24}\
\ee^{4\pi i\m u J_0^3}\m\right).
\label{cb}
\qqq
Here \s$\phi^{l'}_{j\m l}(z): \hat{V}_{k,l}\longrightarrow
\hat{V}_{k,l'}\otimes V_j\s$ are the primary fields of the model
\cite{TK} mapping the unitary highest weight irreducible modules
$\hat{V}_{k,l}$ of the affine Kac-Moody algebra $\hat{sl}_2$
of level $k=1,2,\dots\m$ and spin
$l=0,{1\over2},\dots,{k\over 2}\m\s$.
\m$V_j$ stands for the spin $j$ module of $sl_2$ and
the conformal blocks take values in the vector space
$\otimes_nV_{j_n}$.
\m$L_0$ is the Virasoro generator and $J_0^3$ the Cartan
subalgebra generator of $sl_2$ acting in $\hat{V}_{k,l}$.
$q$ is shorthand for $\ee^{2\pi i\tau}$.
\vskip 0.3cm

The elliptic conformal blocks $\gamma_{\tau,\un z}(u)$
given by Eq.\s\s\pref{cb} satisfy a second order differential
equation \cite{Ber} generalizing the genus zero
Knizhnik-Zamolodchikov (KZ) equations \cite{KZ}. In the absence
of insertions, the genus one blocks coincide with the characters
of the $\hat{sl}_2$ modules $\hat{V}_{k,l}$. For the case
with insertions, free field realizations of modules
$\hat{V}_{k,l}$ and of the primary fields \cite{Wak}\cite{bf}
allow to write integral representations for the genus one
conformal blocks. An original approach
was developed in \cite{Etin0}\cite{Etin} where the elliptic
conformal blocks were identified with vector-valued
functions on the Kac-Moody group covariant w.r.t. the adjoint
action of the loop group.
\vskip 0.3cm

The genus one correlation functions
of the WZW conformal field theory
are sesqui-linear combinations of its conformal blocks.
Geometrically, the conformal blocks may be viewed as
holomorphic sections of (Friedan-Shenker \cite{FShen})
vector bundles over the Teichm\"{u}ller space of punctured
elliptic curves parametrized by $(\tau,\un z)$. It is
a hermitian structure on the Friedan-Shenker bundle which
determines the correlation functions.
\vskip 0.3cm

There are various possible approaches to the construction
of the Friedan-Shenker bundles for the WZW theory.
In an algebraic approach, one considers
the spaces of invariants
(or coinvariants \cite{UTY}) arising by localization
of irreducible representations of affine
algebra $\hat{sl}_2$ at punctures of the complex curve.
Alternatively, in an algebraic-geometric approach,
one studies spaces of holomorphic
sections of powers of the determinant bundle over
the moduli space of $G$-bundles
with parabolic structure at the punctures.
Finally, in an approach inspired by the intrinsic relation
between the two-dimensional conformal WZW field theory,
and the three-dimensional topological Chern-Simons (CS)
theory observed in \cite{witt}, one considers spaces
of quantum states of the group $G$ CS theory in
the presence of Wilson lines.
The above constructions are believed (and at least partially
proven) to result in equivalent objects but
different approaches lead to different insights into
their nature and permit to use different technical tools
in their analysis. For example, the Kac-Moody approach allows
to apply the representation theory
of the affine algebras and the determinant bundle approach
the powerful tools of modular algebraic geometry.
For genus one, the algebraic approach was recently
extensively discussed (for general simple group) in
\cite{FeW}.
\vskip 0.3cm

One of the virtues of the approach via the CS theory is
that the spaces of quantum states come with a formal
scalar product which should provide the Friedan-Shenker
bundles with a hermitian structure determining
the correlation functions of the WZW theory.
The scalar product of the CS states
is given by a functional integral
which, although not Gaussian, may be calculated
by iterative Gaussian integration. This leads to
finite-dimensional integral expressions
obtained in \cite{Quadr} (in the $SU_2$ case) and in \cite{fgk}
(in the case of general simple group) for genus zero and
in \cite{g} and \cite{g1}  for
genus $>\s 1$ (in the $SU_2$ case). For genus one, only the
case with no insertions and general simple group
was treated up to now in \cite{gk} where it was shown
that characters of the unitary representations of the
Kac-Moody algebra are
orthonormal with respect to the scalar product of
the CS states. We partially fill this gap here by deriving
the integral expressions for the scalar product of
arbitrary elliptic CS
states for the $SU_2$ group. For the case with a single
Wilson line, we prove the convergence of the integrals
and the compatibility of the scalar product with the
Knizhnik-Zamolodchikov-Bernard (KZB) connection defining
the differential equations satisfied by the conformal
blocks. We also relate our integral expressions to those
for conformal blocks derived by the free field techniques
\cite{bf}. For the genus zero case, the latter were
used to obtain the Bethe Ansatz solutions for simple
spin chains \cite{babfl}\cite{reshvar}
and this relation has been recently extended
to the elliptic case \cite{FeVar}.
\vskip 0.4cm

The paper is organized as follows. In Sect.\s\s2, we recall
the description of the elliptic CS states for the $SU_2$
gauge group worked out in \cite{fg}. In Sect.\s\s3, we
perform the iterative Gaussian integral obtaining a
finite-dimensional integral expression for the scalar
product. In Sect.\s\s3 we analyze the one-point
case proving the unitarity of the KZB connection
with respect to the scalar product of one-point states.
Sect.\s\s4 briefly illustrates the relation of our approach
to the contour integral representations of conformal blocks
and to the Bethe Ansatz on the example of
elliptic one-point blocks. This discussion recaps the
results of \cite{Etin-1}. Finally, in Section 5, we
explain in what sense our results provide an exact
solution of the genus one WZW theory.
\vskip 0.9cm

\nsection{Elliptic CS states}
\vskip 0.5cm

In \cite{fg} we have described the space of states of
the $SU_2$ Chern-Simons theory
on  $T^2\times\NR$ ($T^2$ is the two dimensional torus
and $\NR$ is the time axis) at level $k$ and in the presence
of Wilson lines in
representations of spin $j_n$ $n=1,\dots,N$.
We shall recall below the construction of this space.
\vs 0.3cm

A complex structure given by the coordinate
$z=\sigma_1+\tau\sigma_2$ on $T^2$
(\s$\equiv\s\{(\sigma_1,\sigma_2)\in\NR^2/(2\pi\NZ^2)\}$),
where $\tau\equiv\tau_1+i\tau_2$ with $\tau_2>0$, induces
a complex structure on the space
of (smooth) $SU_2$ connections on the torus
and with the use of
the latter one may carry out the Bargmann quantization
of the theory.
States $\Psi$ are holomorphic functionals of
the $0,1$-component of the connection
$A^{01}\equiv A_{\bar z}d\bar z$. They take values
in the tensor product $\otimes_n V_{j_n}$ of the
representation spaces of $SU_2$.
Smooth maps $g:T^2\rightarrow SL_2$
act on $A^{01}$ by complex gauge transformations
\qq
A^{01}\ \mapsto\ {}^g\hskip -1mm A^{01}=gA^{01}g^{-1}+g\de g^{-1},
\nonumber
\qqq
with $\de=d\bar z\partial_{\bar z}$.
States corresponding to Wilson lines
$\{z_n\}\times\NR$ should verify the chiral Ward identity
\qq
\Psi(^g\hskip -1mm A^{01})=\ee^{kS(g^{-1},A^{01})}
\otimes_n g(z_n)_{_{(n)}}\Psi(A^{01})\label{chwi}
\qqq
where $S(g,A^{01})$ is the action of the
Wess-Zumino-Witten model coupled to
$A^{01}$, see \cite{fg}, and the subsubscript
$(n)$ indicates that the $SL_2$ element acts
on the space $V_{j_n}$ of spin $j_n$ representation.
\vs 0.3cm

Restricting $\Psi$ to connections
$A^{01}_u=-u\sigma_{3} d\bar z/(2\tau_2)$,
$u\in\NC$, we can assign to every
state a
holomorphic map
$\gamma:\NC\rightarrow\otimes_nV_{j_n}$
related to $\Psi$ by the equation
\qq
\Psi(A_u^{01})= \ee^{\pi k u^2/\tau_2}
\otimes_n (\ee^{\sigma_3u(\bar z_n-z_n)/(2\tau_2)})_{_{(n)}}
\gamma(u).\label{gamma}
\qqq
It was shown in \cite{fg}
that the correspondence between holomorphic maps
$\gamma$ and states $\Psi$ is one to one
provided that $\gamma$ satisfies the following conditions
\qq
\gamma(u+1)&=&\gamma(u),\alabel{cond}a\cr
\gamma(u+\tau)&=&\ee^{-2\pi ik(\tau+2u)}
\otimes_n(\ee^{iz_n\sigma_3})_{_{(n)}}\gamma(u),\aeqno b\cr
0&=&\oplus_n(\sigma_3)_{(n)}\gamma(u),\aeqno c\cr
\gamma(-u)&=&\otimes_n (w)_n\gamma(u),\aeqno d\cr
\qqq
and
\qq
&&\partial_u^{l_0}
\partial_{v_1}^{l_1}\dots
\partial_{v_N}^{l_{_N}}
\otimes_n\hs{-0.07cm}
\left(\langle j_n\vert\,(\ee^{v_n\sigma_{+}})_{_{(n)}}
\right)\gamma(u)
\vert_{_{u=\alpha\hfill\atop v_n=\exp[(\alpha-\bar\alpha)z_n/\tau_2]}}=0
\clabel{reg}\cr\cr
&&\hbox{for every $N+1$-tuple of non negative integers with}
\ \sum_{n=0}^N l_n<\sum_{n=1}^Nj_n\cr\cr
&&
\hbox{and $\alpha=0,
\,{1\over2},\,{\tau\over2},\,{\tau+1\over2}$}.
\qqq
$w$ in (\ref{cond},d) is the generator of the Weyl
group of $SL_2$ and $\vert j_n\rangle$
is the highest weight vector
of the representation of spin $j_n$ i.\s e.
$(\sigma_{3})_{_{(n)}}\vert j_n\rangle=
2j_n\,\vert j_n\rangle$. We shall denote the space
of maps $\gamma(u)$ satisfying the above conditions
by $W_{\tau,\un z,\un j}$. The combined results
of \cite{UTY}\cite{Etin} show that
this space is isomorphic to the space of
Kac-Moody invariants localized on the elliptic
curve\footnote{we thank G. Felder for this remark}.

\vs 0.9cm

\nsection{Scalar product of CS states}
\vs 0.5cm

We shall be interested in a study of the scalar product
of Chern-Simons states. The latter will be expressed
as a finite dimensional integral (the actual
dimension will depend on the spin of the insertions)
which, if finite, will provide $W_{\tau,\un z,\un j}$
with a natural structure of a Hilbert space.
The finiteness of the integral has been proven only in
particular cases.
Formally the scalar product in  holomorphic
quantization is given by the functional integral
\qq
\parallel \Psi\parallel^2=
\int \vert\Psi(A^{01})\vert^2\s\s
\ee^{{ik\over 2\pi}\int\tr(A^{10} A^{01})}
\ DA\label{sp}
\qqq
where $A^{10}=A_zdz$ with $A_z=-A_{\bar z}^*$.
Our strategy for the calculation of the integral
\pref{sp} will be as in \cite{gk}.
\vskip 0.5cm

\subsection{Change of variables}
\vskip 0.3cm

\noindent The first observation is that an open dense set
of connections $A^{01}$ can be
parametrized by $^{g^{-1}}\hskip-1.5mm A_u$
with $g$ a chiral gauge
transformation and
$u$ in some fundamental domain
of the action of translations
$u\rightarrow u+1 $ and $u\rightarrow u+\tau$
and reflection $u\rightarrow -u$ on
$\NC$.
With this restriction, the only ambiguity in the
parametrization of $A^{01}$ is left multiplication of $g$
by a constant gauge transformation with values in the
Cartan subgroup $T^\NC$.
We shall see below how to treat this freedom.
\vs 0.3cm

Using Eq.\s\s\pref{chwi} we obtain, see \cite{gk}
\qq
&&\vert\Psi(A^{01})\vert^2\s\s
\ee^{{ik\over 2\pi}\int\tr(A^{10} A^{01})}\cr\cr
&=&\langle\Psi(A_u)\s,\s
\otimes_n \left((g g^*)(z_n)\right)^{-1}_{(n)}\Psi(A_u)\rangle\
\ee^{kS(\zeta_u g g^* \zeta_u^*)-\pi k(u^2+\bar u^2)/\tau_2}
\label{unint}
\qqq
with $\zeta_u=\ee^{\sigma_{3}u(z-\bar z)/(2\tau_2)}$.
Note that the right hand side of Eq.\s\s\pref{unint} depends
only on the product
$gg^*$. It is then natural to use the Iwasawa decomposition
for $g\in SL_2$,
\qq
g=bU\label{iwa1}
\qqq
with $U\in SU_2$ and
\qq
b=\ee^{\sigma_{+}v}\ee^{\sigma_{3} \varphi/2},\label{iwa2}
\qqq
for $v\in\NC$ and ${\varphi\in\NR}$.
We may write Eq.\s\s\pref{sp} in the new variables as
\qq
\parallel \Psi\parallel^2&
=&\int \langle\Psi(A_u)\s,\s
\otimes_n \left((g g^*)(z_n)\right)^{-1}_{(n)}\Psi(A_u)\rangle
\ j(u,g)\ \delta(\varphi(0))
\ \cr\
&&\hspace{0.4cm}
\times\s\s\ee^{kS(\zeta_u g g^* \zeta_u^*)-\pi
k(u^2+\bar u^2)/\tau_2}\ j(u,g)\ \delta(\varphi(0)
\ Dg\ d^2u,
\label{newvar}\qqq
with $Dg=\Pi_z dg(z)$ the formal product of invariant measures
on $SL_2$.
The Jacobian from $A$ variables to $g$ and $u$ has
been computed in \cite{gk}:
\qq
j(u,g)=C\,\ee^{4S(\zeta_ugg^*\zeta_u^{*})}\,
\m\vert\Pi(u,\tau)\vert^4
\nonumber\qqq
with $C$ a constant independent of $u$ and $\tau$
and
\qq
\Pi(u,\tau)=q^{1/8}\sin(2\pi u)\prod_{r=1}^\infty
(1-q^r)(1-\ee^{4\pi iu}q^r)(1-\ee^{-4\pi iu}q^r),
\qqq
The $\delta$-function in
\pref{newvar} fixes partially the ambiguity
in the parametrization of $A^{01}$, the rest of the
the freedom (left multiplication of $g$
by a constant gauge transformation
with values in the Cartan subgroup $T$ of $SU_2$) is harmless:
as the integrand is constant and $T$ is compact
it only introduces an irrelevant normalization factor.
The invariant measure of $SL_2$
in the parametrization of Eqs.\s\s\pref{iwa1} and \pref{iwa2}
is
\qq
dg= d\varphi\ d^2(e^\varphi v)\ dU
\nonumber\qqq
with $dU$ the Haar measure of $SU_2$.
The integrand of \pref{newvar} does not
depend on the $U$ variables.
Their integration will give a factor independent
of the other variables
that we absorbe in the normalization.
The action with this parametrization is given by
\qq
S(\zeta_u gg^*\zeta_u^{*})=-{i\over 2\pi}\int(\da\varphi)(\de\varphi)
-{i\over 2\pi}\int\ee^{-2\varphi_\chi}(\da\bar v_u)(\de v_u)
+{\pi\over \tau_2}(u-\bar u)^2
\nonumber\qqq
where $\chi=(z-\bar z)u/(2\tau_2)$, $v_u=v \ee^{(z-\bar z)u/\tau_2}$
and $\varphi_\chi=\varphi+\chi+\bar\chi$.
Notice that although $v$ is a function on the torus,
$v_u$ is not but it has a monodromy
in the $\tau$ direction depending on $u$
\qq
v_u(z+2\pi)=v_u(z)\qquad v_u(z+2\pi\tau)=
\ee^{4\pi i u}v_u(z).\label{monod}
\qqq
Writing $(gg^*)^{-1}$ in terms of variables $v_u$ and $\varphi$,
\qq
(gg^*)^{-1}=
\ee^{-\bar v\sigma_{-}}
\ee^{-\varphi\sigma_{3}}
\ee^{-v\sigma_{+}}
=\zeta_u^*
\ee^{-\bar v_u\sigma_{-}}
\ee^{-\varphi_\chi\sigma_{3}}
\ee^{-v_u\sigma_{+}}
\zeta_u
\nonumber\qqq
and using the map $\gamma(u)$
introduced in relation \pref{gamma},
we obtain the following
expression for the scalar product:

\qq
\parallel\Psi\parallel^2\ =&&
C\int\ee^{-{i(k+4)\over 2\pi}
[\int(\da\varphi)(\de\varphi)
+\int\ee^{-2\varphi_\chi}(\da\bar v_u)(\de v_u)
+2i\pi^2(u-\bar u)^2/\tau_2]}\cr
&&\hskip -1cm\times\langle\gamma(u)\s,\s\otimes_n\left(
\ee^{-\bar v_u\sigma_{-}}
\ee^{-\varphi_\chi\sigma_{3}}
\ee^{-v_u\sigma_{+}}\right)(z_n)_{_{(n)}}\gamma(u)\rangle
\ D\varphi\, D(e^\varphi v)\,d^2u\s.
\label{sp2}
\qqq
\vs 0.4cm

\subsection{$v$-integral}
\vs 0.3cm

\noindent Now we must perform the functional integrals.
First notice that the $v$-integral is Gaussian so it can be
easily obtained using the two point function
\qq
&&\int \ v_u(z)\, \bar v_u(z')
\ \ee^{-i(k+4)(2\pi)^{-1}\hskip -1mm\int\ee^{-2\varphi_\chi}
(\da\bar v_u)(\de v_u)}\ D(e^\varphi v)\clabel{twop}\cr\cr
&=&{\pi\over k+4} N(u,\tau)\
\ee^{i\pi^{-1}[\int(\da\varphi)(\de\varphi)
+2i\pi^2(u-\bar u)^2/\tau_2]}
\int \ee^{2\varphi_\chi(y)}\, P_u(z-y)\, \overline{P_u(z'-y)}\
d^2y\s,
\qqq
where
\qq
N(u,\tau)=
(q\bar q)^{1/24} \,\vert\Pi(u,\tau)\vert^{-2}
\prod_{n=1}^\infty\vert 1-q^n\vert^2
\nonumber\qqq
and
\qq
P_u(z)=-{\vartheta_1'(0)\,\vartheta_1(z-4 \pi u)\over
\pi\,\vartheta_1(z)\,\vartheta_1(4\pi u)}
\qqq
is the Green function of \m$\de$ on functions with monodromies
given by \pref{monod}. Here $\vartheta_1$ is the
theta-function s.t.
\qq
\vartheta_1(z+2\pi)=-\vartheta_1(z)\qquad
\vartheta_1(z+2\pi\tau)=-\ee^{-i(z+\pi\tau)}\vartheta_1(z).
\qqq
It is odd, has simple zeros for
$z\in 2\pi\NZ+2\pi\tau\NZ$ and is non zero elsewhere.
We shall take $\vartheta_1(z)=2\Pi({z\over 4\pi},\tau)$.
\s$P_u(\m\cdot\m)$ is a meromorphic function for
$u\not\in(\NZ+\tau\NZ)/2$ as one could expect
from the fact that in those cases the line bundle defined
by conditions \pref{monod}
has no holomorphic sections
different from zero.
\vs 0.3cm

Now let us insert into the matrix element
$\langle\gamma(u),\cdots\gamma(u)\rangle$
a complete basis of
$\otimes_n V_{j_n}$ composed of eigenvectors
of every $(\sigma_{3})_{(n)}$ with eigenvalues
$2(p_n-j_n)$
and let us denote these vectors by $\vert\ \un p\ \rangle$
with $\un p\equiv(p_1,\dots,p_N)$. Then
\qq
&&\langle\gamma(u)\s,\s\otimes_n\left(\ee^{-\bar v_u\sigma_{-}}
\ee^{-\varphi_\chi\sigma_{3}}
\ee^{-v_u\sigma_{+}}\right)(z_n)_{_{(n)}}\gamma(u)\rangle\cr\cr
&=&
\sum_{\un  l\atop 0\leq p_n\leq 2j_n}\prod_n\ee^{2(j_n-p_n)
\varphi_\chi(z_n)}
|\, \langle\ \un p\ \vert
\otimes_n\left(\ee^{-v_u\sigma_{+}}\right)(z_n)_{_{(n)}}\gamma(u)\rangle
\,|^2.\label{part}
\qqq
We may perform the $v$-integral in Eq.\s\s\pref{sp2}
with the use of relations \pref{twop} and \pref{part} obtaining
\qq
&&\hs{-0.2cm}\parallel\Psi\parallel^2\ =\ C
\sum_{\un p}\left({\pi\over k+4}\right)^K{1\over K!}
\int N(u,\tau)\ \vert \Pi(u,\tau)\vert^4
\cr\cr &&\times\s|\sum_{\un p'\atop |\un p'|=K}
F(\un z,\un y,\un p')\
\langle\ \un p\ \vert \otimes_n\left((p'_n!)^{-1}
(\sigma_{+})_{(n)}^{p'_n}\right)\gamma(u)\rangle
\ |^2\s\clabel{sp3}\cr\cr
&&\times \int \prod_{s=1}^K\ee^{2\varphi_\chi(y_s)}
\prod_{n=1}^N\ee^{2(j_n-p_n)\varphi_\chi(z_n)}
\ \ee^{{-i(k+2)\over2\pi}[\int (\da\varphi)(\de\varphi)
+2i\pi^2(u-\bar u)^2/\tau_2]}\s
\delta(\varphi(0))\s D\varphi\s d^{2K}\un y\s d^2u\s.
\qqq
The condition
$$\sum_{n=1}^N p'_n\equiv|\un p'|=K\equiv\sum_{n=1}^N (p_n-j_n)$$
in the second sum of \pref{sp3} is a consequence
of the invariance of $\gamma(u)$ under the diagonal action
of the Cartan subgroup.
\qq
F(\un z,\un y,\un p')=\sum_{\rho\in S_K}
\prod_{s=1}^K P_u(z^{{\un p'}}_s-y_{\rho(s)})
\qqq
where $S_K$ is the symmetric group,
$\un y \equiv(y_1,\dots,y_K)$
and
$\un z^{{\un p'}}\equiv(z_1,\dots,z_1,z_2,\dots,z_2,\dots,z_N)$
whith every $z_n$ repeated $p'_n$ times.
\vs 0.5cm

\subsection{$\varphi$-integral}
\vs 0.3cm

We still have to evaluate the $\varphi$ integral,
\qq
\int \prod_{s=1}^K\ee^{2\varphi_\chi(y_s)}
\prod_{n=1}^N\ee^{2(j_n-p_n)\varphi_\chi(z_n)}
\ \ee^{-i{k+2\over2\pi}\int( \da\varphi)(\de\varphi)}
\ \delta(\varphi(0))\ D\varphi
\label{phint}
\qqq
corresponding to a Coulomb gas system of zero total
charge. As noted before,
the neutrality is a consequence of the invariance
of $\gamma(u)$ under the diagonal Cartan subgroup.
The Green function $G(z-z')$ of the operator
$\partial_z\partial_{\bar z}$ restricted to
functions on the torus orthogonal to constants is
\qq
G(z)={\ln\vert\vartheta_1(z)\vert^2\over\pi}+
{(z-\bar z)^2\over8\pi^2\tau_2}+ {\rm const}.
\qqq
The Gaussian integral \pref{phint} is equal to
\qq
C (q\bar q)^{-1/24}
\tau_2^{-1/2}\prod_{n=1}^\infty\vert 1-q^n\vert^{-2}
&&\hskip -3mm
\prod_s\ee^{2(\chi(y_s)+\bar\chi(y_s))}
\prod_n\ee^{2(j_n-p_n)(\chi(z_n)+\bar\chi(z_n))}\cr\clabel{phint2}
\cr
&&\times\s\s\ee^{-{\pi\over 4(k+2)}\int f(z)G(z-z')f(z')}
\qqq
with
\qq
f=\sum_n 2(j_n-p_n)\delta_{z_n}+\sum_s 2\delta_{y_s}.
\nonumber\qqq
The Green function  behaves as $\pi^{-1}\ln{\vert z-z'\vert^2}$ when
$z\rightarrow z'$ and, consequently,
expression \pref{phint2} diverges due to the terms diagonal
in the $\delta$-functions of $f$. These divergences
may be regularized by splitting the coinciding points
to distance $\epsilon$ and
removed by multiplicative
renormalization of the scalar product.
The divergence is proportional to
\qq
\epsilon^{-{2\over k+2}(\sum_n (p_n-j_n)^2+\sum_n(p_n - j_n))}\s.
\nonumber\qqq
It is the most severe for $p_n=2j_n$ since $0\leq p_n\leq 2j_n$.
We shall then renormalize the point-split version
of \pref{phint2} by multiplication by
$\epsilon^{{2\over k+2}\sum_n j_n(j_n+1)}$
and taking $\epsilon$ to zero.
The effect of the
renormalization is to remove
all terms in the sum over $\un p$  in Eq.\s\s\pref{sp3}
except those with the most severe divergence, i. e. $p_n=2j_n$.
\vs 0.3cm

Combining expressions \pref{sp3} and \pref{phint2}, we obtain,
after the renormalization,
\qq
\parallel\Psi\parallel^2\s=\s
C\,\tau_2^{-1/2}\int
\ee^{(\pi(k+2)\tau_2)^{-1}(\Theta-\overline\Theta)^2}
\ \vert \omega(\un z,\un y,u,\gamma)\vert^2\label{spf}
\qqq
with
\qq
\Theta=
\pi(k+2)u+{_1\over^2}\sum_sy_s-{_1\over^2}\sum_n j_n z_n
\label{Th}
\qqq
and $\omega$ the multivalued form
\qq
\omega(\un z,\un y,u,\gamma)&=&\vartheta'_1(0)^{-{1\over k+2}\sum_n
j_n(j_n+1)}
\prod_{n<n'}\vartheta_1(z_n-z_{n'})^{-2j_nj_{n'}/(k+2)}
\prod_{s,n}\vartheta_1(y_s-z_n)^{2j_n/(k+2)}
\cr\cr&&
\times\prod_{s<s'}\vartheta_1(y_s-y_{s'})^{-2/(k+2)}\
\Pi(u,\tau)\ dy_{_1}\wedge\cdots\wedge dy_{_J}\wedge du
\clabel{omega}\cr\cr
&&\times\sum_{\un p\atop
|\un p|= J}F(\un z,\un y,\un p)\
\otimes_n\left(\langle j_n \vert
(p_n!)^{-1}({\sigma_{+}})_{(n)}^{p_n}\right)\gamma(u)\rangle\s,
\qqq
where $J=\sum_nj_n$. Note that $|\omega(\un z,\un y,u,\gamma)|^2$
is univalued in $z_n,
y_s,2\pi u\in T^2$. Eq.\s\s\pref{spf} expresses the scalar
product of current
blocks as a finite dimensional
integral in variables
$2\pi u$ and $y_s$ over $(T^2)^{1+J}$.
\vskip 0.8cm

\nsection{Convergence of integrals}
\vskip 0.5cm

The expression under the integral
in \pref{spf} has
apparent non integrable divergencies.
It is expected that singularities
cancel or at least become integrable
when $\gamma$ satisfies conditions \pref{cond} and \pref{reg}.
For instance, we shall see below that, although
$\omega$ in \pref{omega}
seems to diverge at $u=0,\,1/2,\,\tau/2,\,(\tau+1)/2$,
it actually depends on $u$ in a holomorphic way.
\vs 0.3cm

Before going to the general case let us study
a simpler one with a unique insertion. This case may be
easily controlled.
Suppose that the insertion at $z$ has spin
$j$ (necessarily $j$ is an integer).
We shall denote  by $\gamma_j$ the vector valued holomorphic
map associated to the state $\Psi_j$ by equation \pref{gamma}.
Function
\qq
\theta(u)\equiv\langle j\vert(\sigma_{+})^j \gamma_j(u)\rangle\
\Pi(u,\tau)
\label{thet}
\qqq
is a theta-function of degree $2(k+2)$ satisfying
\qq
\quad\theta(u+1)=\theta(u),
\quad\theta(u+\tau)=\ee^{-2\pi i(k+2)\m(\tau+2u)}\s\theta(u)\s.
\label{per}
\qqq
An explicit basis of such functions is formed by
expressions
\qq
\theta_n(u,\tau)\s=\s\sum\limits_{r=-\infty}^\infty
q^{\m(k+2)({n\over 2(k+2)}+r)^2}\s\s\ee^{\m 4\pi i u(k+2)
({n\over 2(k+2)}+r)}
\label{bas}
\qqq
with $n=0,1,\dots,2k+1$. \m$\theta_n$ satisfy the heat equation
\qq
(\da_\tau+{i\over
8\pi(k+2)}\m\da_u^{\m2})\s\s\theta_n\s=\s0\s.
\label{heat}
\qqq
Besides, the functions $\theta$ coming from Eq.\s\s\pref{thet}
have to have a definite parity
\qq
\theta(-u)=(-1)^{j+1}\m\theta(u)
\nonumber\qqq
and have to obey the selection rules \cite{fg}
\qq
\partial_u^l\theta_j(u)\vert_{u=\alpha}=0\qquad\hbox{for}
\ l\leq j\ \hbox{and}\
\alpha=0,\,{\textstyle {1\over2},\,{\tau\over2},\,{\tau+1\over2}}.
\label{reg1}
\qqq
That leaves a $(k+1-2j)$-dimensional space of solutions.
Labeling states $\gamma_j$ with one insertion
by the theta-functions $\theta(u)$, we obtain
from Eq.\s\s\pref{omega}
\qq
\omega((y_s-z),u,\theta)\ \s=\ \s
{_{j!}\over^{pi^j}}\hspace{1mm}\theta(u)\
\prod\limits_{s=1}^j P_u(z-y_s)\s
\prod_{s<s'}\tilde\vartheta_1(y_s-y_{s'})^{-2/(k+2)}\s\cr
\times\s\prod_{s=1}^j\tilde\vartheta_1(z-y_s)^{2j/(k+2)}\
dy_1\wedge\cdots\wedge dy_j\wedge du\s\cr
={_{j!}\over^{\pi^j}}\hspace{1mm}\theta(u)\
\tilde\vartheta_1(4\pi u)^{-j}\s
\prod_{s<s'}\tilde\vartheta_1(y_s-y_{s'})^{-2/(k+2)}\s
\prod_{s=1}^j\tilde\vartheta_1(z-y_s-4\pi u)\s\cr
\times\s\prod_{s=1}^j\tilde\vartheta_1(z-y_s)^{-1+2j/(k+2)}\
dy_1\wedge\cdots\wedge dy_j\wedge du\s,
\label{ome}
\qqq
where $\tilde\vartheta_1(y)\equiv\vartheta_1(y)/\vartheta_1'(0)$.
The $\un y$ integral in Eq.\s\s\pref{spf} has a positive
degree of divergence when $y_s\to z$ (recall that
$\tilde\vartheta_1(y)\sim y$ when $y\to 0$) and it may
be easily shown to converge. As for the $u$-dependence,
the singular behavior $\sim(u-\alpha)^{-j}$ around $u=\alpha$
coming from $\tilde\vartheta_1(4\pi u)^j$
is regularized due to the conditions
\pref{reg1} for $\theta_j$ which guarantee that
$\omega$ is holomorphic in $u$. We infer the
convergence of the integral on the right hand side
of Eq.\s\s\pref{spf}:
\vs 0.1cm

{\bf the one-point states are
normalizable.}
\vs 0.1cm

\no Note that due
to the presence of $\Pi(u,\tau)$ vanishing at $u=\alpha$
in the expression for $\omega$, the normalizability
would still hold if we replaced the condition
$l<j$ in \pref{reg1} by $l<j-1$ which, however, given the
parity properties of the states, implies the stronger
condition. Hence, for the one-point states,
the normalizability is equivalent to the fusion
conditions \pref{reg1}.
We expect this to hold for general states.
\vs 0.3cm

Let us return to the general case
and let us study the dependence of \m$\omega$ on $u$.
We only have to examine the
last line of Eq.\s\s\pref{omega} since the
possible divergencies in $u$
come from the zeros of $\vartheta_1(u)$
in the denominator of $P_u(z)$ in $F$.
Consider first the case $u\hskip-1mm\rightarrow\hskip-1mm0$
and write
\qq
P_u(z_n-y_s)=-{1\over4\pi^2 u}+f_{n,s}(u)
\nonumber\qqq
with $f_{n,s}$ analytic
around $u=0$. Let us examine
for fixed $\un p$
the coefficient in $F(\un z,\un y,\un p)$ of the term
\qq
\left({-1\over4\pi^2 u}\right)^{J-L}
f_{1,1}f_{1,2}\cdots f_{1,l_1}
f_{2,l_1+1}\cdots f_{2,l_1+l_2}\cdots
f_{N,L}
\nonumber\qqq
where $L=\sum_n l_n$. Denote this coefficient by $c_{\un p,\un l}$,
and call $C_{\un l}$ similar coefficient arising after the sum
over $\un p$ in the last line of Eq.\s\s\pref{omega}.
With a little combinatorics one easily obtains
\qq
c_{\un p,\un l}=(J-L)!\hspace{1mm}{p_1!\cdots p_N!\over
(p_1-l_1)!\cdots (p_N-l_N)!}
\nonumber\qqq
and adding up all contributions from different
$\un p's$ one has
\qq
C_{\un l}&=&
(J-L)!\sum_{\un p\atop|\un p|= J}
\otimes_n\left(\langle j_n \vert
((p_n-l_n)!)^{-1}({\sigma_{+}})_{(n)}^{p_n}\right)\gamma(u)\rangle\cr
&=&(J-L)!\hspace{2mm}
\partial_{v_1}^{l_1}\dots
\partial_{v_N}^{l_N}
\otimes_n\left(\langle j_n\vert\,(\ee^{v_n\sigma_{+}})_{_{(n)}}
\right)\gamma(u)
\vert_{_{v_n=1}}.
\nonumber\qqq
{}From condition \pref{reg} we know that the last expression
vanishes at least as
$u^{J-L}$ when $u\rightarrow 0$.
Identical computation
shows regularity of \m$\omega$ at $u=1/2$ and a slight
modification of it allows to deal with
cases $u=\tau/2,\, (1+\tau)/2$. The final conclusion
is that $\omega$ depends holomorphically
on $u$ provided that conditions
\pref{reg} are satisfied. That in the general case the
singularities in $y$'s become integrables remains
still to be proven.
\vskip 0.9cm

\nsection{Unitarity of KZB connection}
\vskip 0.5cm

The spaces of Chern-Simons states $W_{\tau,\un z,\un j}$ form a
holomorphic vector bundle $W_{\un j}$
over the space $\NC\times(\NC^N\setminus
\Delta_N)$, where $\Delta_N$ contains the $N$-tuples $(z_1,\dots,z_N)$
with coincidences modulo $2\pi(\NZ+\tau\NZ)$.
A (smooth) family of scalar products on $W_{\tau,\un z,\un j}$
provides bundle $W_{\un j}$ with a hermitian structure.
On the other hand, $W_{\un j}$ may be equipped
with a holomorphic Knizhnik-Zamolodchikov-Bernard (KZB)
connection first described in the elliptic case
in \cite{Ber}, see
\cite{fg} for its presentation
using the same formalism as the one employed here.
It is related to elliptic versions of integrable models
\cite{Etin}\cite{FeW}.
It has been conjectured in \cite{Quadr}\cite{Karp} that
the KZ connection and its higher
genus generalizations (see \cite{Ber2}\cite{Hitch}\cite{AdPW})
are unitary i.e. that there exists a hermitian structure
preserved by the parallel transport and that, moreover,
such a hermitian structure is provided by the rigorous version of
the formal scalar products \pref{sp} of Chern-Simons states.
It is a simple fact that a hermitian structure determines
a unique holomorphic unitary connection. This way,
the hermitian structure preserved by the KZ connection,
if existent, may be regarded as a more basic object
that the connection itself. The conjecture was proven
for special insertions
for genus zero. The main obstruction in proving it for
general insertions was the lack
of control of the convergence of the
finite-dimensional integral expressions to which the
functional integral in \pref{sp} was reduced
for genus zero in \cite{Quadr} and for genera $\geq\s 2$
in \cite{g}\cite{g1}. For genus one and no insertions,
the conjecture follows from the orthogonality of the
Kac-Moody characters in the scalar product \pref{spf},
established in \cite{gk}, and from the well known fact
that the characters satisfy a heat equation
on the moduli space. Below, we shall provide a further
support for the conjecture by proving it for genus one
case with one insertion.
\vskip 0.3cm

For a single insertion of spin
$j$ at point $z$, the KZB connection
is given by the formulae \cite{fg}
\qq
\nabla_{\bar z}\s\theta(u)&=&\da_{\bar z}
\s\s\theta(u)\s,\cr\cr
\nabla_z\s\theta(u)&=&\da_z\s\s\theta(u)\s,\cr\cr
\nabla_{\bar\tau}\s\theta(u)&=&\da_{\bar\tau}
\s\s\theta(u)\s,\cr\cr
\nabla_{\tau}\s\theta(u)&=&\left(\da_{\tau}\s+
\s{_i\over^{8\pi(k+2)}}
\s\da_u^{\m2}\s-\s{_{\pi i j(j+1)}\over^{2(k+2)}}\s
\da_u\m P_u(y)\m\vert_{_{y=0}}\right)\s\theta(u)\s.
\label{KZB}
\qqq
The potential
\qq
\da_u\m P_u(y)\m\vert_{_{y=0}}\s=\s4\s\CP(4\pi u)\s-\s
8\sum\limits_{n=1}^\infty{_{q^n}\over^{(1-q^n)^{2}}}
\s+\s{_1\over^3}\s,
\label{Wei}
\qqq
where $\CP(y)$ is the Weierstrass function (meromorphic function
on $T^2$ with the only pole at $0$ where $\CP(y)=y^{-2}+\CO(y^2)$).
Hence the $\tau$-component of the KZB connection contains
the Lam\'{e} operator $\da_u^2+c\s\CP(4\pi u)$ (which is replaced
by the elliptic Calogero-Sutherland operator for the
$SU_n$ group \cite{Etin}).
\vskip 0.3cm

The integral \pref{spf} giving the scalar product of the states
has a natural geometric interpretation which we shall
spell out for the one-point case. $(j+1)$-form
$\omega(\un y,u,\theta)$ may be viewed
as a holomorphic form on $T^2\times((T^2)^j\setminus\Delta_j)$
with values in a complex line bundle $\CL\s$ ($\Delta_j$ contains
$j$-tuples $\un y$ with $y_s=y_{s'}$
for some $s\not=s'$ or with $y_s=0$
for some $s$). By definition, the sections $\sigma$ of $\CL$
have the same transformation properties under
$y_s\mapsto y_s+2\pi,\m2\pi\tau$, $u\mapsto u+1,\m\tau$
and under the pure braiding of $y_s$ as $\omega$.
The univalued expression
$$\ee^{(\pi(k+2)\tau_2)^{-1}(\Theta-\overline\Theta)^2}\s\s
\vert\sigma(\un y,u)\vert^2$$ defines a hermitian structure
on $\CL$ which, in turn, determines a unique unitary
connection
\qq
\nabla_{\bar y_s}\sigma&=&\da_{\bar y_s}\sigma\s,\cr
\nabla_{y_s}\sigma&=&\left(\da_{y_s}\m+\m{_1\over^{
\pi(k+2)\tau_2}}\s(\Theta-\bar\Theta)\right)\sigma\s,\cr
\nabla_{\bar u}\m\sigma\s\m &=&\da_{\bar u}\s\sigma\s,\cr
\nabla_{u}\sigma\s\s &=&\left(\da_{u}\m+\m{_2\over^{
\tau_2}}\s(\Theta-\bar\Theta)\right)\sigma\s.
\qqq
The connection acts naturally on differential forms
with values in $\CL$ and its unitarity implies that,
if $\eta,\s \chi$ are two $\CL$-valued forms
of degree $j+1$ and $j$, respectively,
and $\eta$ is holomorphic then
\qq
d\left(\ee^{(\pi(k+2)\tau_2)^{-1}(\Theta-\overline\Theta)^2}
\s\s\overline{\eta}\s\s\s\chi\right)\s=\s
\ee^{(\pi(k+2)\tau_2)^{-1}(\Theta-\overline\Theta)^2}
\s\overline{\eta}\s\s\s\nabla\chi\s.
\label{d1}
\qqq
Consequently,
\qq
\int\ee^{(\pi(k+2)\tau_2)^{-1}(\Theta-\overline\Theta)^2}\s
\overline{\eta}\s\s\s\nabla\chi\s=\s0
\label{dd}
\qqq
if the integral converges and there are no
contributions from coinciding $y_s$ or $y_s=0$ to the integral
of the left hand side of Eq.\s\s\pref{d1}.
This observation will be at the core of an argument
showing that the KZB connection is unitary.
\vs 0.3cm

The unitarity of the KZB connection with respect
to the hermitian structure defined by the one-point
version of the formula \pref{spf} is equivalent to the
following relations:
\qq
{_d\over^{dz}}\s\parallel\Psi\parallel^2&=&
C\,\tau_2^{-1/2}\int
\ee^{(\pi(k+2)\tau_2)^{-1}(\Theta-\overline\Theta)^2}
\ \overline{\omega(\un y,u,\theta)}\s\s\s
\omega(\un y,u,\m\da_z\m\theta)\s,\cr
{_d\over^{d\tau}}\s\parallel\Psi\parallel^2&=&
C\,\tau_2^{-1/2}\int
\ee^{(\pi(k+2)\tau_2)^{-1}(\Theta-\overline\Theta)^2}
\ \overline{\omega(\un y,u,\theta)}\s\s\s
\omega(\un y,u,\nabla_\tau\theta)\s,
\label{Uni}
\qqq
where we assumed that \s$\theta$ depends
holomorphically on $z$ and $\tau$. $\Theta$ is given
by Eq.\s\s\pref{Th} with $z$ set to zero. The first
of these equations is obvious since, after the shift
$y_s\mapsto y_s+z$, the $z$-dependence
under the integral appears only in $\theta$.
The proof of the second relation is more involved because
the $\tau$-dependence occurs both under the integral
and in the integration domain parametrized by
$y_s,\m 2\pi u\in 2\pi(\NC/(\NZ+\tau\NZ))$.
Reparametrizing $y_s$ as $r_s+\tau t_s$ and $2\pi u$
as $\chi+\tau\phi$, we obtain by differentiation
under the integral
\qq
{_d\over^{d\tau}}\s\parallel\Psi\parallel^2\s=\s
C\,\tau_2^{-1/2}\int
\ee^{(\pi(k+2)\tau_2)^{-1}(\Theta-\overline\Theta)^2}
\ \overline{\omega(\un y,u,\theta)}\s\s\hs{4cm}\cr
\times\bigg(\da_\tau\s+\s{_{u-\bar u}\over^{2i\tau_2}}\s\da_u
\s+\s\sum\limits_{s=1}^j\da_{y_s}\s{_{y_s-\bar y_s}\over^{2i\tau_2}}
\s-\s{_{i}\over^{2\pi(k+2)\tau_2^2}}\s(\Theta-\bar\Theta)^2\m
-\s{_i\over^{4\tau_2}}\bigg)\s\omega(\un y,u,\theta)\s.
\label{afin}
\qqq
The differentiation under the integral is easily
substantiated since it does not spoil the integrability.
Writing
\qq
\omega(\un y,u,\theta)\s\equiv\s\theta(u)
\s\s\tilde\omega(\un y,u)\s,
\nonumber\qqq
see Eq.\s\s\pref{ome}, we obtain after a straightforward
but somewhat tedious algebra:
\qq
\bigg(\da_\tau\s+\s{_{u-\bar u}\over^{2i\tau_2}}\s\da_u
\s+\s\sum\limits_{s=1}^j\da_{y_s}\s{_{y_s-\bar y_s}\over^{2i\tau_2}}
\s-\s{_{i}\over^{2\pi(k+2)\tau_2^2}}\s(\Theta-\bar\Theta)^2\m
-\s{_i\over^{4\tau_2}}\bigg)\s\omega(\un y,u,\theta)\s\cr\cr
=\ (\nabla_\tau\theta(u))\s\s\tilde\omega(\un y,u)
\s+\s\nabla\chi(\un y,u)\s+\s\theta(u)\s\s\psi(\un y,u)\s,
\qqq
where the $j$-form $\chi$ and the $(j+1)$-form $\psi$
with values in the bundle $\CL$
are given by the following explicit expressions:
\qq
&&\chi(\un y,u)\cr\cr
&&=\s{_{(-1)^ji}\over^{8\pi(k+2)}}\bigg(\theta(u)
\s\s(\s\da_u+\sum_s{_{y_s-\bar y_s}\over^{\tau_2}}\s)\s\s\m
\hat\omega(\un y,u)\s
-\m\bigg((\da_u+{_{2\pi(k+2)}\over^{\tau_2}}
\m(u-\bar u)\m)\s\theta(u)\bigg)\s\hat\omega(\un y,u)
\bigg)\cr\cr
&&+\m\sum\limits_{s=1}^j{_{(-1)^{s}i}
\over^2}\m\theta(u)\hs{-0.01cm}\prod\limits_{s'\not=s}
P_u(-y_{s'})\s(\da_u+{_{y_s-\bar y_s}\over^{\tau_2}})\s P_u(-y_s)
\s R_j(\un y)^{2\over k+2}\m dy_1\wedge
\smash{\mathop{\cdots}\limits_{\hat{s}}}
\wedge dy_j\wedge du\s,\hs{0.6cm}\\ \cr\cr
&&\psi(\un y,u)\ =\ \bigg(\da_\tau\s-\s{_i\over^{8\pi(k+2)}}
\s\da_u^{\m 2}\s+\s{_{\pi ij(j+1)}\over^{2(k+2)}}\s\da_u P_u(y)
\vert_{_{y=0}}\bigg)\s\tilde\omega(\un y,u)\s\cr
&&\hs{1cm}+\s{_i\over^2}\sum\limits_{s=1}^j\da_{y_{s}}\hs{-0.05cm}
\bigg(
(\da_u P_u(-y_{s}))\prod\limits_{s'\not=s}P_u(-y_{s'})\s\s
R_j(\un y)^{2\over k+2}\bigg)\s dy_1\wedge\dots\wedge dy_j\wedge du
\label{ps}
\qqq
in which $\hat\omega$ stands for $\tilde\omega$ without the
$du$ differential and
\qq
R_j(\un y)\s=\s\prod\limits_{s<s'}\tilde\vartheta_1(y_s-y_{s'})^{-1}
\prod\limits_{s=1}^j\tilde\vartheta_1(y_s)^j\s.
\qqq
\vskip 0.2cm

First note that \s$\nabla\chi(\un y,u)$
does not contribute to the integral on
the right hand side of Eq.\s\s\pref{afin} due to the
relation \pref{dd}. This, in fact, requires some analysis.
We should prove that the corresponding terms are
integrable and do not lead to boundary contributions. As for
the $u$-dependence, the form $\chi$ is still holomorphic
in $u$. Although the $u$-derivatives increase the degrees
of the poles at $u=\alpha$ by one, the fusion rules \pref{reg1}
still guarantee the regularity. As for the $y_s$-behavior,
note that, although the $y_s$-derivatives produce additional
$\sim y_s^{-1}$ factors, they act on forms with an improved
$y_s$ behavior \s(\s$\da_u P_u(-y_s)$ is regular at $y_s=0$\s)\m.
\m The same argument is used to show that there are no
boundary contributions from $y_s=0$. The additional
$\sim(y_s-y_{s'})^{-1}$ factors which may be produced by the
$y_s$-differentiation also do not spoil the
integrability and there are no boundary contributions
from $y_s=y_{s'}$ (in showing this
one uses the fact that such terms
may be present only for $k>4$ and that, due to the fusion rule,
$j<{k\over 2}$). We leave the details to the reader.
\vskip 0.3cm

In order to complete the proof of relation \pref{Uni} and of
the unitarity of the KZB connection for one-point insertions,
we shall show that $\psi(\un y,u)$, given by Eq.\s\s\pref{ps},
vanishes. We have
\qq
\psi(\un y,u)\s\s=\s\bigg(\sum\limits_{s=1}^j\m
(\s(\da_\tau+{_i\over^2}
\da_{y_s}\da_u)\m P_u(-y_s)\s)\prod\limits_{s'\not=s}
P_u(-y_{s'})\s\s\cr\cr
+\ {_2\over^{k+2}}\s\m U_j(\un y,u)\bigg)\m
R_j(\un y)^{2\over{k+2}}\m\s dy_1\wedge\dots\wedge dy_j
\wedge du\s,
\nonumber\qqq
where
\qq
U_j(\un y,u)&\equiv&\prod\limits_{s=1}^j P_u(-y_s)\s\s
\da_\tau \ln{R_j(\un y)}
\s-\s{_i\over^{16\pi}}\sum\limits_{s=1}^j
(\s\da_u^{\m 2} P_u(-y_s)\s)\prod
\limits_{s'\not=s} P_u(-y_{s'})\s\cr
&-&{_i\over^{8\pi}}\sum\limits_{s<s'}
(\s\da_u P_u(-y_s)\s)\s(\s\da_u P_u(-y_{s'})\s)
\prod\limits_{s''\not=s,s'} P_u(-y_{s''})\s\cr
&+&{_i\over^{2}}\sum\limits_{s=1}^j
(\s\da_u^{\m 2} P_u(-y_s)\s)\prod
\limits_{s'\not=s} P_u(-y_{s'})\s\da_{y_s}
\ln{R_j(\un y)}\s\cr
&+&{_{\pi ij(j+1)}\over^{4}}
(\s\da_u P_u(y)|_{_{y=0}}\s)\prod
\limits_{s=1}^j P_u(-y_{s})\s.
\label{uj}
\qqq
$F(y)\equiv(\da_\tau+{_i\over^2}\da_{y}\da_u)\m P_u(-y)$ is
a holomorphic function of $y$ such that
\qq
F(y+2\pi)=F(y)\s,\hs{1cm}F(y+2\pi\tau)=\ee^{-4\pi iu}\s F(y)\s.
\nonumber\qqq
Hence it vanishes. Similarly,
\qq
U_j((y_s'+2\pi\m\delta_{s\m s'}),u)\s\ &=&U_j(\un y,u)\s,\cr
U_j((y_s'+2\pi\tau\m\delta_{s\m s'}),u)&=
&\ee^{-4\pi iu}\s\s U_j(\un y,u)\s\s.
\nonumber\qqq
$U_j(\un y,u)$ is, {\it a priori}, a meromorphic
function of $\m y_j$ with simple poles possible
at $y_j=y_s,\ s<j,$  and at $y_j=0$. In fact, it is
easy to see that the residues at $y_j=y_s$ vanish
and that the residue at $y_j=0\m$ is equal
to $-{1\over\pi}\s U_{j-1}
((y_s)_{_{s<j}},u)$. Since $U_0=0$, it follows by
induction that all $U_j$ vanish. This ends the proof
of unitarity of the one-point KZB connection.
\vs 0.9cm

\nsection{Integral representations of
conformal blocks and Bethe Ansatz}
\vskip 0.5cm

It was remarked in \cite{Quadr}\cite{gg}
that the integral expressions
for the scalar products of the genus zero Chern-Simons
states are very closely related to the integral expressions
\cite{bf}\cite{VScht} for
the genus zero conformal blocks, i.e. for the sections
horizontal w.r.t. the KZ connection. The integral
expressions for the conformal blocks are the consequence of
the Wakimoto (free field) realization of the
current algebra \cite{Wak} and they are
given by contour integrals of the genus zero
counterpart of the forms $\omega$ with $k+2$ replaced
by $-(k+2)$. Similarly, one may use the
results obtained above to write down integral expressions
for the genus one one-point conformal blocks, see also
\cite{bf}\cite{Etin-1}. Namely, consider
a vector of holomorphic $j$-forms with components
\qq
&\Omega_n(\un y,u)\ =\ \theta_n(\m u-{_1\over^{2\pi(k+2)}}
\sum_s\m y_s)
\s\ \hat\omega'(\un y,u)&
\qqq
where the theta-functions $\theta_n$ are given by
Eq.\s\s\pref{bas} and
\qq
\hat\omega'(\un y, u)\s=\s{_{j!}\over^{\pi^j}}\prod\limits_{s=1}^j
P_u(-y_s)\s\s R_j(\un y)^{-{2\over k+2}}\s\m dy_1\wedge\dots
\wedge dy_j\s.
\nonumber\qqq
Note that $\hat\omega'$ is same
as $\hat\omega$ used in Sect.\s\s 5 except for the flip
of sign of $k+2$. In its dependence on
$\un y$, \s$\Omega$ may be viewed as
a holomorphic $j$-form on $(T^2)^j\setminus\Delta_j$
with values in a flat vector bundle. The one-point
genus one conformal blocks may be expressed by the
integrals
\qq
\int\limits_c\Omega_n(\m\cdot\m,u)
\qqq
of the components of $\Omega$ over cycles $c$ of a homology
with values in the dual bundle, see \cite{bf}\cite{FeW}.
The horizontality of such integrals
(in their dependence on the modular parameter $\tau$)
w.r.t. the KZ connection follows from the relation
\qq
&&\bigg(\da_\tau\s+\s{_i\over^{8\pi(k+2)}}\m\da_u^{\m2}\s-\s
{_{\pi i j(j+1)}\over^{2(k+2)}}\s\da_u P_u(y)\vert_{_{y=0}}\bigg)
\s\Omega_n(\un y,u)\s\cr
&&=\s{_1\over^{2i}}\sum\limits_{s=1}^j
\da_{y_{s}}\hs{-0.05cm}\bigg(\theta_n(u)\s\m
(\da_u P_u(-y_{s}))\prod\limits_{s'\not=s}P_u(-y_{s'})\s\s
R_j(\un y)^{2\over k+2}\bigg)
dy_1\wedge\dots\wedge dy_j
\nonumber\qqq
which is an immediate consequence of Eq.\s\s\pref{heat}
and of the vanishing of the form
$\psi'$ given by Eq.\s\s{\pref{ps}} with the sign of $k+2$
inverted.
\vskip 0.3cm

Papers \cite{bab}\cite{babfl}\cite{reshvar}
have remarked that the integral expressions for
the genus zero conformal blocks give rise, in the limit
$k\to -2$ ($k\to -$ dual Coxeter number,
for general simple groups) to Bethe Ansatz solutions
for the Gaudin spin chains, see also \cite{ffr}.
This relation has its
elliptic counterpart \cite{Etin-1}\cite{FeVar}. In order
to illustrate it on the example of the one-point
functions, following \cite{Etin-1}, let us note that the identity
$U_j(\un y,u)\equiv 0$ with $U_j$ given by Eq.\s\s\pref{uj},
implies that
\qq
\left(\da_u^{\m2}-4\pi^2j(j+1)\s\da_u P_u(y)|_{_{y=0}}\right)
\prod\limits_{s=1}^j P_u(-y_s)\ =\
-16\pi i\s\m\da_\tau\ln{R_j(\un y)}\
\prod\limits_{s=1}^j P_u(-y_s)
\label{eiva}
\qqq
if $\un y$ is a stationary point of \m$\ln{R_j}$, i.\s e. if
\qq
j\s{_{\vartheta'_1(y_s)}\over^{\vartheta_1(y_s)}}
\s=\s\sum\limits_{s'\not= s}
{_{\vartheta'_1(y_s-y_{s'})}\over^{\vartheta_1(y_s-y_{s'})}}
\label{BA}
\qqq
for $s=1,\dots,j$. Eq.\s\s\pref{eiva} is the eigenvalue
equation for the Lam\'{e} operator
\s$L\s\equiv\s\da_u^{\m2}-4\pi^2j(j+1)\s\da_u P_u(y)|_{_{y=0}}\s$
or \s$L'\s\equiv\s\da_u^{\m2}-16\pi^2j(j+1)\s\CP(4\pi u)\s$\m.
\m It states that the Bethe Ansatz meromorphic "multi-particle"
wave function
\qq
\psi_{\un y}(u)\s=\s\prod\limits_{s=1}^j P_u(-y_s),
\nonumber\qqq
built of one-particle functions $u\mapsto P_u(-y_s)$,
is an eigenfunction of $L$ or $L'$ with eigenvalue
\qq
\lambda_{\un y}\s=\s-16\pi i\m\s\da_\tau\ln{R_j(\un y)}
\nonumber\qqq
or \s$\lambda'_{\un y}=\m\lambda_{\un y}\m
+\m32\pi^2j(j+1)(\sum\limits_{n=1}^\infty
{_{q^n}\over^{(1-q^n)^{2}}}\m-\m{_1\over^{24}}\m)\m$,
\m respectively, if $y_s$ satisfy Eq.\s\s\pref{BA}. Note that
\qq
\psi_{\un y}\s\cong\s({_{-1}\over^{4\pi^2u}})^j
\hs{0.5cm}{\rm as}\hs{0.5cm}u\to 0
\label{p1}
\qqq
and that
\qq
\psi_{\un y}(u+{_1\over^2})\s=\s\psi_{\un y}(u)\s,
\hs{0.7cm}\psi_{\un y}(u+{_\tau\over^2})\s=\s\ee^{-i\sum_s
y_s}\s\m\psi_{\un y}(u)
\label{p2}
\qqq
so that the eigenfunction $\psi_{\un y}$ corresponds
to the quasi-momenta $0$ and $\sum_s y_s^s$ along the two
cycles of the torus. The above relations reduce the calculation
of the eigenfunction of the Lam\'{e} operator $L$
with properties \pref{p1} and \pref{p2} to solving
the Bethe Ansatz equations \pref{BA}.
\vskip 0.9cm

\nsection{Conclusions}
\vskip 0.5cm

We have reduced in Sect.\s\s 3 the formal functional-integral
formula for the scalar product of
elliptic CS states to finite-dimensional
integral expressions. Under the assumption of
convergence of the integrals, checked only for
states with no or with a single Wilson line,
our result determines a Hilbert
space structure on the space of CS states. This structure
allows, in turn, to define the genus one correlation
functions of the "diagonal" ($A$-series, in the
terminology of \cite{CIZ}) $SU_2$
WZW model in the external field
$A=-(A_u^{01})^*+A_u^{01}$. They are
given by the expressions
\qq
\sum\limits_{\alpha}\s\ee^{\m\pi k(u-\bar u)^2/\tau_2}\ \s
(\s\otimes_n\ee^{\sigma_3 u(\bar z_n-z_n)/
(2\tau_2)}\s\s\gamma_\alpha(u)\s)
\s\otimes\s(\s\overline{
\otimes_n\ee^{\sigma_3 u(\bar z_n-z_n)/
(2\tau_2)}\s\s\gamma_\alpha(u)}\s)\s,
\label{Cf}
\qqq
where $(\gamma_\alpha)$ corresponds through Eq.\s\s\pref{gamma}
to an orthonormal basis of the elliptic CS states.
The correlation functions for the generic unitary gauge field
$A=-(A^{01})^*+A^{01}$ with $A^{01}=\s{}^{g^{-1}}\hs{-0.23cm}A^{01}_u$
are obtained by acting on \pref{Cf} by
\qq
\ee^{\m k S(gg^*,\s-(A_u^{01})^*+A_u^{01})}\s\
(\s\otimes_n g(z_n)^{-1}\s)\s\otimes\s(\s
\overline{\otimes_n g(z_n)^{-1}}\s)\s.
\nonumber\qqq
The above expressions are independent of the choice
of the orthonormal basis $(\gamma_\alpha)$ of states
and may be computed using an arbitrary basis by replacing
\qq
\sum\limits_\alpha\m\gamma_\alpha\otimes\overline{\gamma_\alpha}
\hs{0.6cm}
{\rm by}\hs{0.6cm}\sum_{\alpha,\beta}\m
H^{\beta\alpha}\s\gamma_\alpha\otimes
\overline{\gamma_\beta}\s,
\nonumber\qqq
where the matrix $(H^{\alpha\beta})$ is the inverse of the
matrix of scalar products
$H_{\alpha\beta}=(\gamma_\alpha,\gamma_\beta)$.
Hence the calculation of the correlation functions reduces to
the computation of the scalar products of elliptic CS states.
The latter are given by (the polarized version of) Eq.\s\s\pref{spf}
and require calculation of the finite-dimensional integrals.
This way, our result provides an exact solution
for the elliptic correlation functions of the
diagonal $SU_2$ WZW model. For the non-diagonal
$D$-series version of the model which corresponds to
the $SO_3$ WZW theory, the correlation functions
are given by similar formulae but with non-overlined
$\gamma_\alpha$ replaced by
\qq
\hf\s \sum\limits_\rho\s{}^\rho\gamma_\alpha
\nonumber\qqq
where $\rho$ runs through the group $Hom(\pi_1(T^2
\setminus\{z_1,\dots,z_N\})\s,\s\s\NZ_2)$ which
acts projectively and unitarily
on the space of elliptic CS states,
see \cite{gg} for the discussion of the
case without insertions. In both diagonal and
non-diagonal WZW theories, the scalar product
of CS states is the essential structure which
allows to determine the correlation functions.
It is ultimately related to the KZ connection
and its higher genus generalizations
as well as to the Bethe Ansatz. It certainly
deserves further study.

\end{document}